\documentclass[10pt,journal]{IEEEtran}
\pdfoutput=1
\usepackage{graphicx}
\usepackage[justification=justified]{caption}
\usepackage{ragged2e}
\usepackage{enumerate}
\usepackage{url}
\usepackage{multirow}
\usepackage{amsmath}
\usepackage{bm}
\usepackage{hyperref}
\usepackage{threeparttable} 
\usepackage{booktabs} 
\usepackage{amsmath,bm} 

\usepackage{subfigure}
\usepackage{float}
\usepackage{makecell}
\usepackage{cite}

\begin{document}

\title{Deniable Steganography}

\author{
	Yong~Xu,~
	Zhihua~Xia,
	Zichi~Wang,
	Xinpeng~Zhang,
	and~Jian~Weng
	\thanks{Yong Xu is with Engineering Research Center of Digital Forensics, Ministry of Education, Nanjing University of Information Science \& Technology, Nanjing, 210044, China.}%
	\thanks{Zhihua Xia and Jian Weng are with College of Cyber Security, Jinan University, Guangzhou, 510632, China. Zhihua Xia is also with Engineering Research Center of Digital Forensics, Ministry of Education, Nanjing University of Information Science and Technology, Nanjing, 210044, China.}%
	\thanks{Zichi Wang and Xinpeng Zhang are with School of Communication and Information Engineering, Shanghai University, Shanghai, 200444, China.}%
	\thanks{Zhihua Xia is the corresponding author. E-mail: xia\_zhihua@163.com.}}


\maketitle

\begin{abstract}
	Steganography conceals the secret message into the cover media, generating a stego media which can be transmitted on public channels without drawing suspicion. As its countermeasure, steganalysis mainly aims to detect whether the secret message is hidden in a given media. Although the steganography techniques are improving constantly, the sophisticated steganalysis can always break a known steganographic method to some extent. With a stego media discovered, the adversary could find out the sender or receiver and coerce them to disclose the secret message, which we name as coercive attack in this paper. Inspired by the idea of deniable encryption, we build up the concepts of deniable steganography for the first time and discuss the feasible constructions for it. As an example, we propose a receiver-deniable steganographic scheme to deal with the receiver-side coercive attack using deep neural networks (DNN). Specifically, besides the real secret message, a piece of fake message is also embedded into the cover. On the receiver side, the real message can be extracted with an extraction module; while once the receiver has to surrender a piece of secret message under coercive attack, he can extract the fake message to deceive the adversary with another extraction module. Experiments demonstrate the scalability and sensitivity of the DNN-based receiver-deniable steganographic scheme.
\end{abstract}

\begin{IEEEkeywords}
	Steganography, deniability, deep neural networks, coercive attack
\end{IEEEkeywords}

\section{Introduction} \label{sec:introduction}

\begin{figure}[h]
	\centering
	\subfigure[]{\includegraphics[width=.5\textwidth]{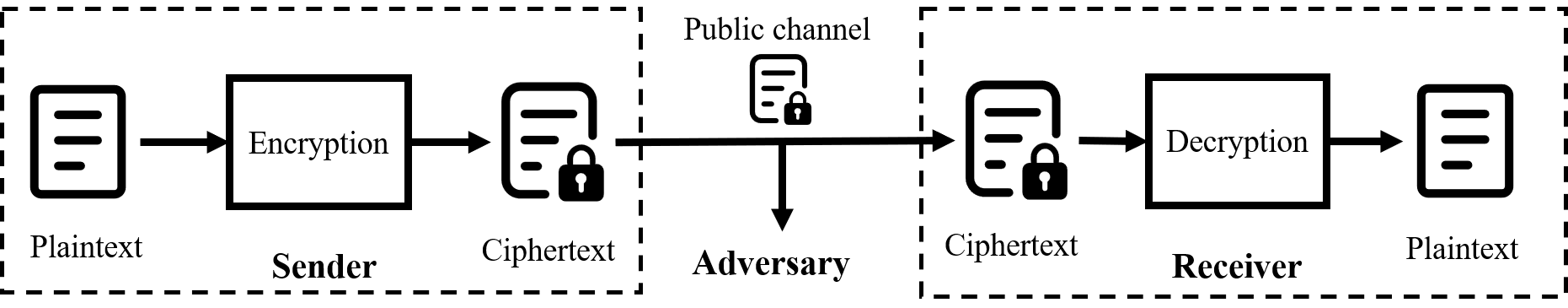}}\\
	\subfigure[]{\includegraphics[width=.5\textwidth]{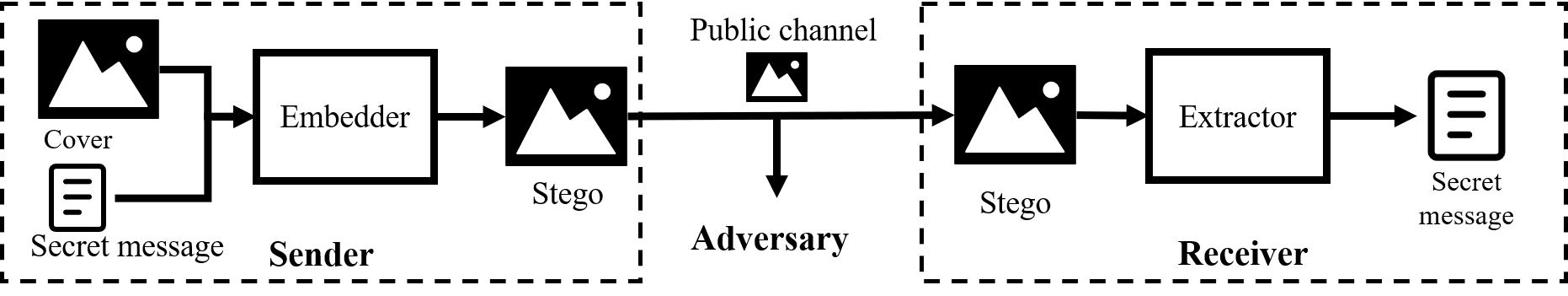}}
	\centering
	\caption{The framework of secure communication in a public channel by (a) cryptography and (b) steganography. The communication parties, i.e., the sender and receiver, could be coerced by the adversary to disclose the secret message in some situations.}
	\label{fig:FrameworkOfCommunication}
\end{figure}

 \IEEEPARstart{I}{nformation} security has been faced with threats and attacks from ancient spy warfare to nowadays data disclosure. To protect the security of communication, cryptography and steganography are two commonly used techniques. As illustrated in Fig.~\ref{fig:FrameworkOfCommunication}, the former encrypts the message to perturb its content, while the latter conceals the message in a cover to hide the existence of the message itself \cite{petitcolas1999information, cheddad2010digital}. However, both of them could encounter \textit{coercive attack}. For cryptography, coercive attack certainly happens since the ciphertext is always suspicious to the adversaries. Specifically, the adversary could coerce the communication parties (the sender or receiver) to disclose the secret messages and keys. To verify the disclosed message, the adversary may further ask the sender to generate the ciphertext again with the message and secret key. As to steganography, the adversary can also coerce the communication parties to disclose the secret message when the existence of the secret message has been detected. In such a malicious scenario, is it still possible to guarantee the information security?

\begin{figure}[h]
	\centering
	\subfigure[]{\includegraphics[width=.48\linewidth]{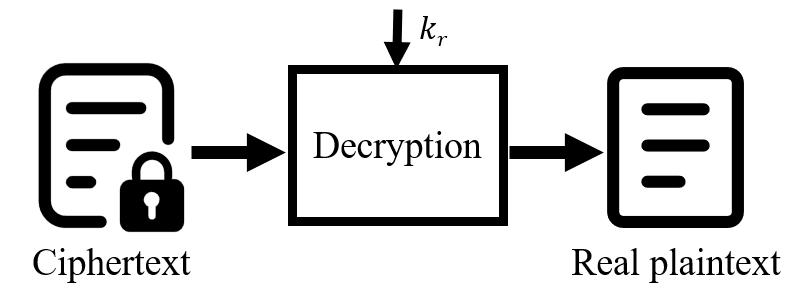}}
	\subfigure[]{\includegraphics[width=.48\linewidth]{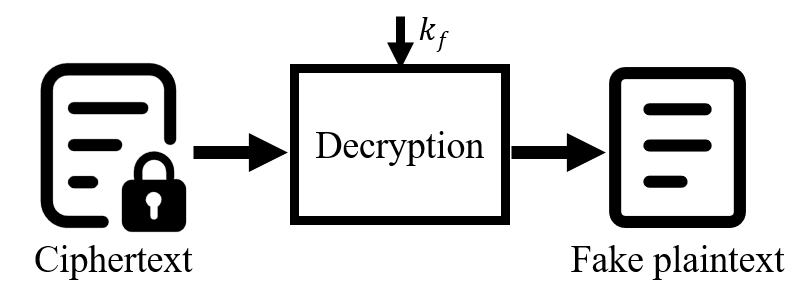}}
	\centering
	\caption{A sketch for the receiver-deniable encryption scheme, (a) a real plaintext will be obtained when the real key is used, and (b) a fake plaintext can be obtained when the fake key is used. When the receiver is coerced, he can surrender the fake plaintext to deceive the adversary.}
	\label{fig:Sketch for receiver-deniable encryption}
\end{figure} 

 In cryptography, deniable encryption has been proposed to deal with \textit{coercive attack} \cite{canetti1997deniable, klonowski2008practical, ibrahim2009method, ibrahim2009receiver, o2011bi, moldovyan2014bi, durmuth2011deniable, sahai2014use, li2017deniable, chi2021not}. As an example in Fig. \ref{fig:Sketch for receiver-deniable encryption}, in a receiver-deniable scheme, the coerced receiver can decrypt the ciphertext with a fake key to obtain a fake message that looks like a real message. However, current researches about steganography mainly focus on the imperceptibility and capacity. The deniability against coercive attack has never been considered yet. On the other hand, many excellent steganalysis methods have been continuously proposed to detect the existence of hidden messages, which are great threats to steganography. For instance, SRM \cite{fridrich2012rich}, GNCNN \cite{qian2015deep}, and SRNet \cite{boroumand2018deep} can detect STC-based steganographic algorithms (e.g., HUGO \cite{pevny2010using}, WOW \cite{holub2012designing}, S-UNIWARD \cite{holub2014universal}) with high accuracy from 70\% to 80\%, and ATS \cite{lerch2016unsupervised} can detect the DNN-based steganography methods (e.g., HiDDeN \cite{zhu2018hidden}) with the accuracy up to 98\% if the model weights are known. It increases the probability of coercive attack on steganography.

\begin{figure}
	\centering
	\subfigure[]{\includegraphics[width=.48\linewidth]{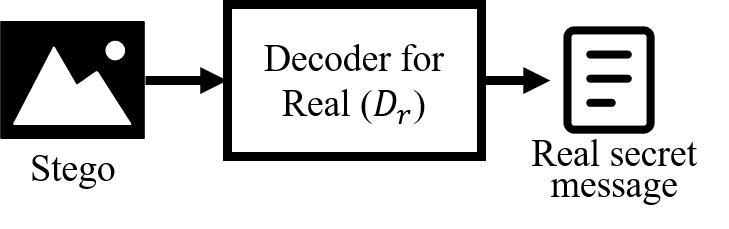}}
	\subfigure[]{\includegraphics[width=.48\linewidth]{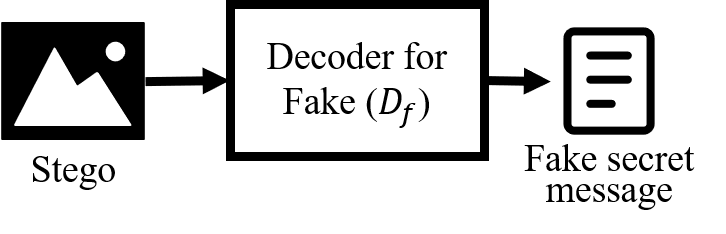}}
	\centering
	\caption{The framework of receiver-deniable steganography, (a) a real secret message can be obtained when $D_r$ is used, but (b) a fake secret message will be obtained when $D_f$ is used. When the communication parties are coerced, he can extract the fake secret message to deceive the adversary.}
	\label{fig:Framework of deniable steganography}
\end{figure} 

 In this paper, for the first time, we consider the deniable steganography. We build up the definitions about deniable steganography according to that in deniable encryption. Then, we construct a receiver-deniable steganographic scheme using deep neural networks (DNN). As shown in Fig. \ref{fig:Framework of deniable steganography}, we design two extractors (named as the \textit{decoders} in DNN-based methods) for the extraction of hidden messages: one for the real message and the other for the fake. Once coerced by the adversary, the receiver can extract and submit a fake message to the adversary but keep the real one private. The contributions of this paper can be concluded as follows:

\begin{itemize}
	\item We build up the concepts about deniable steganography and figure out the requirements, challenges, and possible constructions. To the best of our knowledge, we are the first to consider the deniable steganography. 	
	\item We propose a receiver-deniable steganographic scheme by adjusting the existing DNN-based steganographic methods. The balanced message loss function is designed to balance the extraction accuracy of decoders for real and fake messages. 
	\item We discussed the scalability and sensitivity of our receiver-deniable steganographic scheme by constructing it with different numbers of decoders and different steganographic networks. The results demonstrate that our scheme has good scalability to the number of decoders and is not sensitive to the structures of encoders and decoders. 
\end{itemize}

 The rest of this paper is organized as follows. Related works are presented in Section~\ref{sec:related_work}. Definitions about deniable steganography are introduced in Section~\ref{sec:definition}. We propose a receiver-deniable steganographic scheme in Section~\ref{sec:scheme}. Experimental results and analysis are provided in Section~\ref{sec:experiments}. Section~\ref{sec:conclusion} concludes the whole paper.

\section{Related Work}\label{sec:related_work}

 Some typical methods about the deniable encryption and DNN-based steganography are introduced here.

\subsection{Deniable Encryption}

 The deniable encryption was firstly considered in 1997 by Ran Canetti \textit{et al.} \cite{canetti1997deniable}. The authors mainly discussed the \textit{sender-deniable} encryption scheme as the \textit{receiver-deniable} and \textit{sender-and-receiver-deniable} schemes can be constructed by the \textit{sender-deniable} scheme. The authors built up the formal concepts about deniable encryption in both the public-key and shared-key encryption scenarios. In a sender-deniable public-key encryption scheme, given a plaintext $p$, a local random input $r$, the corresponding ciphertext $c$ is calculated as

	 \begin{equation}	 
		c=E_{k}(p, r),	 
	\end{equation}

 \noindent where $E_{k}$ is the encryption algorithm with a public key $k$. To be deniable, the sender should have a faking algorithm $\phi(\cdot)$ to generate a fake random input $\tilde{r}$ by

	\begin{equation}
		\tilde{r}=\phi(p, r, c),
	\end{equation}	

 \noindent where $\tilde{r}$ will be used as the local random input of another plaintext $\tilde{p}$, so as to generate the same   $c$ as,

	\begin{equation}
		c = E_{k}(\tilde{p}, \tilde{r}).
	\end{equation}

 \noindent In this way, the coerced sender can claim that $c$ was generated by $(\tilde{p}, \tilde{r})$, without revealing $p$. At the first glance, it seems impossible to construct such deniable encryption as typical cryptographic methods require different plaintexts to generate distinct ciphertexts. Creatively Canetti \textit{et al.} constructed a sender-deniable public-key encryption scheme using trapdoor permutation \cite{canetti1997deniable}. The public-key deniable encryption scheme can be directly used as a shared-key deniable encryption scheme. Besides, the XOR operation is a much easier tool to construct the shared-key encryption scheme  \cite{canetti1997deniable}. The core functionality of \textit{receiver-deniable} scheme is illustrated in Fig.~\ref{fig:Sketch for receiver-deniable encryption}. The receiver-deniable scheme can be directly constructed by using a sender-deniable scheme twice while the \textit{sender-and-receiver-deniable} scheme needs the intermediaries for construction \cite{canetti1997deniable}.
 
 In 2008, Klonowski \textit{et al.} \cite{klonowski2008practical} claimed that the adversary will find the application of a deniable scheme after opening sufficient bits, which would make the adversary continually ask the sender to open the real plaintext. Thus, a “slightly” banned message, instead of the original message, could be more convincing than the one generated in \cite{canetti1997deniable}. The authors proposed a nested construction to cope with the worse scenario where the adversary knows deniable encryption is used. Besides, they also combined One-Time-Pad, ElGamal Cryptosystem, and the covert channel with deniable encryption for different scenarios respectively. In 2009, Ibrahim \textit{et al.} \cite{ibrahim2009method} proposed a sender-deniable encryption scheme by the quadratic residuosity and trapdoor permutation to obtain higher deniability and efficiency. In the same year, Ibrahim \textit{et al.} \cite{ibrahim2009receiver} proposed a receiver-deniable encryption scheme by using mediated-RSA \cite{boneh2001method} without using the sender-deniable scheme twice. O'Neill \textit{et al.} \cite{o2011bi} proposed two sender-and-receiver-deniable (bi-deniable) encryption schemes without intermediaries. The schemes work in the so-called “multi-distributional” model, in which the parties run alternative key-generation and encryption algorithms for equivocal communication, but claim under coercion to have run the prescribed algorithms. Moldovyan \textit{et al.} \cite{moldovyan2014bi} considered a worse scenario, i.e., active attack, where the adversary can impersonate the sender or receiver. Accordingly, they proposed a \textit{bi-deniable} public-key encryption scheme associated with a probabilistic encryption algorithm. An authentication procedure is performed to both parties using RSA signatures at the beginning, which provides security against active coercive attack. D{\"u}rmuth \textit{et al.} \cite{durmuth2011deniable} claimed that a sender-deniable encryption scheme could be constructed from any samplable public key bit-encryption scheme. Two constructions are designed by quadratic residuosity and trapdoor permutations, respectively. Sahai \textit{et al.} \cite{sahai2014use} proposed a punctured program technique by the pseudo-random function and pseudo-random generator. The punctured program can apply indistinguishability obfuscation to transform a natural private-key encryption scheme into a public-key encryption scheme, and thus can construct the deniable encryption scheme with the XOR operator. Recently, deniable encryption is even combined with searchable encryption to protect data in the cloud storage scenario \cite{li2017deniable, chi2021not}.

 To the best of our knowledge, the deniable scheme for steganography has not been reported in the literature. In this paper, we define the deniable steganography for the first time and propose a receiver-deniable scheme using deep neural networks.

\subsection{Deep Steganography}

 Deep neural networks have been exploited for image steganography due to their flexibility and adaptability \cite{isac2011study, hussain2020survey}. Typical architectures used for deep steganography include the Encoder-Decoder Architecture (EDA) and Generative Adversarial Networks (GAN) \cite{goodfellow2014generative}.
 
 In EDA, the encoder and decoder are designed to embed and extract the secret messages respectively. The loss functions generally include two parts: the image loss and the secret message loss. DeepStego \cite{baluja2017hiding} hides a secret image within another same-size cover image. Apart from the encoder-decoder network, the authors constructed a preparation network to extract more useful features from the secret image for succinct encoding. Then the encoder takes the cover image and features as the input, generating a stego image; and the decoder extracts the secret image from the stego image. Almost immediately, the authors found that the secret image in \cite{baluja2017hiding} can be revealed by a simple blurring operation on the stego, and proposed two strategies to solve the problem\cite{baluja2019hiding}: one is to shuffle the pixels of the secret image before encoding and the other is to hide two secret images into one cover image. StegNet \cite{wu2018stegnet} is also an image-into-image steganographic scheme. It maintains identical structures for the encoder-decoder but combines the skip connections \cite{he2016deep} and separate convolution \cite{chollet2017xception} to form a new block called SCR for quicker convergence. Rahim \textit{et al.} \cite{rahim2018end} designed a two-branch encoder to extract features from the cover and secret images respectively. Then, the features are concatenated to produce stego images. The EDA steganography schemes can embed a large secret image into the cover. But the undetectability of the secret is not considered carefully. 
 
 Generative Adversarial Networks (GAN) simulate a game between a generator and a discriminator to optimize the both. The competition between steganography and steganalysis is rightly the case in GAN. SGAN \cite{volkhonskiy2020steganographic} uses the deep convolutional GAN (DCGAN) \cite{radford2015unsupervised} to generate secure covers, and then adopts manual steganography algorithms to hide secret messages in generated covers. The network includes three components: a generator to generate cover images, a discriminator to detect whether the image is a nature image or a generated one, and a steganalyzer (another discriminator in fact) to discriminate stegos from covers. Similar to SGAN, SSGAN \cite{shi2017ssgan} utilizes Wasserstein GAN \cite{arjovsky2017wasserstein} instead of DCGAN to generate covers. Besides, GNCNN \cite{qian2015deep} is used for the discriminator and the steganalyzer, which generates covers with better visual quality. The adversarial mechanism can improve the undetectability but the generated cover itself can be suspicious. 

 To solve the above problem, ASDL-GAN \cite{tang2017automatic} adopts GAN to just learn an embedding change probability map for nature covers. The discriminator is designed to discriminate between stegos and covers while the generator is to generate the embedding change probability map which is directly related to the undetectability against the discriminator. The probability map is then further transformed to be a modification map by Ternary Embedding Simulator (TES). Finally, a stego is produced by adding the modification map to the cover. Although ASDL-GAN learns to concentrate the modifications on textual regions of covers, it is still inferior to the manual steganography algorithm S-UNIWARD \cite{holub2014universal} under the detection of steganalyzers like SRM \cite{fridrich2012rich} and XuNet \cite{xu2016structural}. UT-SCA-GAN \cite{yang2018spatial} makes three improvements to ASDL-GAN further. Firstly, Tanh-simulator is introduced instead of TES for less pre-training cost and higher fitting accuracy. Secondly, U-Net \cite{ronneberger2015u} is adopted for the generator to improve the security performance and decrease the training time as well. Thirdly, the selection-channel awareness (SCA) \cite{yang2017steganalysis} is incorporated into the discriminator for better resistance against steganalysis. To embed a large quantity of secrets securely, bach steganography is customized to select suitable covers, allocate the payload to different covers and then embed the secrets by STC-based algorithms. Zhong \textit{et al.} \cite{Zhong2021BatchSV} used GANs for cover selection, payload assignment and modification probability map learning. It had higher security compared with STC-based algorithms.

 \begin{figure*}[!ht]
	\centering
	\includegraphics[width=0.9\linewidth]{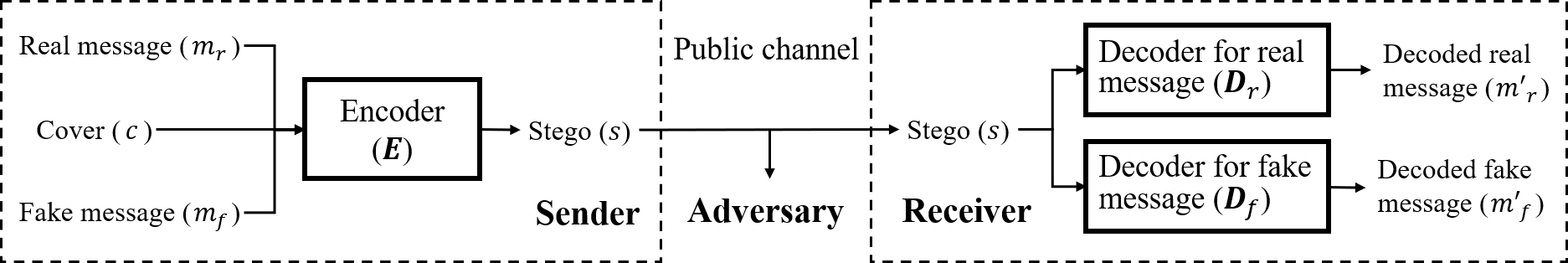}
	\caption{The overall framework of receiver-deniable steganography.}
	\label{fig:scheme_framework}
\end{figure*}

 The embedding map learning \cite{tang2017automatic, yang2018spatial} is not such a direct way to optimize the stegos, and HayesGAN \cite{hayes2017generating} is the first adversarial network to directly generate stegos. It applies a generator to produce stegos with covers and secret messages, a steganalyzer (discriminator) to judge whether an image is a cover or stego, and an extractor to extract messages from stegos. Accordingly, three loss functions: the image loss, the secret message loss, and the adversarial loss, are considered.  HiDDeN \cite{zhu2018hidden} is a hybrid scheme for both steganography and watermarking using GAN. The network architecture and the loss function composition resemble HayesGAN. Additionally, it employs a noise layer to simulate various image distortions for robust watermarking. SteganoGAN \cite{zhang2019steganogan} utilizes more-complex network connections (ResNet \cite{he2016deep} and DenseNet \cite{huang2017densely}) for the generator, so as to improve the image quality and embedding rate. To counteract steganalysis, Peng \textit{et al.} \cite{Peng2022ARC} proposed a coverless steganographic framework based on GANs. The sender uses the secret-mapped nosie to directly produce a stego by the generator. Upon receiving the stego, the receiver tries to approximate the stego by iteration of gradient descent approximation. Finally, the secret can be extracted by inversely maping the receiver's noise.

 Although the GAN-based steganography schemes achieve good undetectability, such schemes still suffer from the newly proposed steganalysis methods. For example, the messages hidden by SGAN \cite{volkhonskiy2020steganographic}, ASDL-GAN \cite{tang2017automatic}, HayesGAN \cite{hayes2017generating}, HiDDeN \cite{zhu2018hidden}, SteganoGAN \cite{zhang2019steganogan} can be detected with high ratios by a new steganalysis method ATS \cite{lerch2016unsupervised}. Then the communication parties will possibly be coerced to reveal the hidden message. 
 
 Aside from the methods mentioned above, new scenarios have sprung up and corresponding steganographic frameworks have been proposed recently. Tao \textit{et al.} \cite{Tao2019TowardsRI} proposed a robust steganographic framework against JPEG compression sicne images are usually lossily compressed through the channels. Firstly, a stego image is obtained by the compressed cover image and the secret. According to the stego image, they adjusted the coefficients of the original cover image to generate an intermediate image, which is identical to the stego image after the same channel compression. As such, by transmitting the intermediate image through the lossy channel, the secret can be extracted correctly. Worse still, it will encounter high risks of detection if the payload increase. To reuse the cover for embedding repeatedly, Wang \textit{et al.} \cite{Wang2022RepeatableDH} proposed a repeatable data hiding framework based on the LSB algorithm. By deducing and designing theoretical modification probabilities, the distortion caused by embedding remains the same even after several times, i.e., the stego is invariable after repeated embedding. While the resistance against statistical analysis has not been valued yet.

\section{Definitions}\label{sec:definition}
\subsection{Basic definitions in steganography}

 The existing steganographic methods can be classified into two categories: manual and DNN-based steganographic methods. A steganography scheme generally includes two algorithms. An embedding algorithm (Encoder) is defined as, 

 	\begin{equation}
 	\label{equ:basicE}
 		s \leftarrow \bm{E} (c, m),
 	\end{equation} 

 \noindent where $c,m,s$ denote the cover image, secret message, and generated stego image, respectively. Then an extraction algorithm (Decoder) is defined as,

  	\begin{equation}
  	\label{equ:basicD}
 		 m' \leftarrow \bm{D} (s),
 	\end{equation}

  \noindent where we have $m'=m$ in manual (non deep-learning) steganographic methods \cite{pevny2010using, holub2012designing, holub2014universal} but may get $m'\approx m$ in most DNN-based methods.
  
\subsection{Definitions about deniable steganography}
 We classify deniable steganography schemes according to which parties are coerced as that in deniable encryption. 
 
 \textit{Sender-deniable steganography} is resilient against coercing of the sender. In this case, the adversary finds a sender sends a stego and coerces the sender to reveal the secret messages. Even more, the adversary may ask the sender to generate the stego again with the original cover and secret message, so as to verify the authenticity of the revealed message. Then, besides the encoder $\bm{E}$ in Eq.~(\ref{equ:basicE}), the sender should have an encoder $\bm{E}_f$ that can generate the same stego $s$ again with a convincing fake message $m_f$ and cover $c_f$ as 

	\begin{equation}
		s \leftarrow \bm{E}_f(c_f,m_f),
	\end{equation} 

 \noindent where $m_f$ should be independent of $m$ but looks like a meaningful message, and $c_f$ is generally similar to $c$.
 
 \textit{Receiver-deniable steganography} is resilient against coercing of the receiver. In this case, the adversary finds a receiver receives a stego and coerces the receiver to extract secret messages. Then, besides the regular $\bm{D}$, the receiver should have a decoder $\bm{D}_f$ that can extract a convincing fake message $m_f$ from the stego as,

	\begin{equation}
		m_f \leftarrow \bm{D}_f(s).
	\end{equation} 

 \textit{Sender-and-receiver deniable steganography} is resilient against coercing of both the sender and the receiver. This can be a simple combination of \textit{sender-deniable} and \textit{receiver-deniable} schemes: the sender has an $\bm{E}_f$ to generate the stego with the fake message and the receiver has a $\bm{D}_f$ to extract the fake message.

 \begin{figure*}[!ht]
	\centering
	\includegraphics[width=0.95\linewidth]{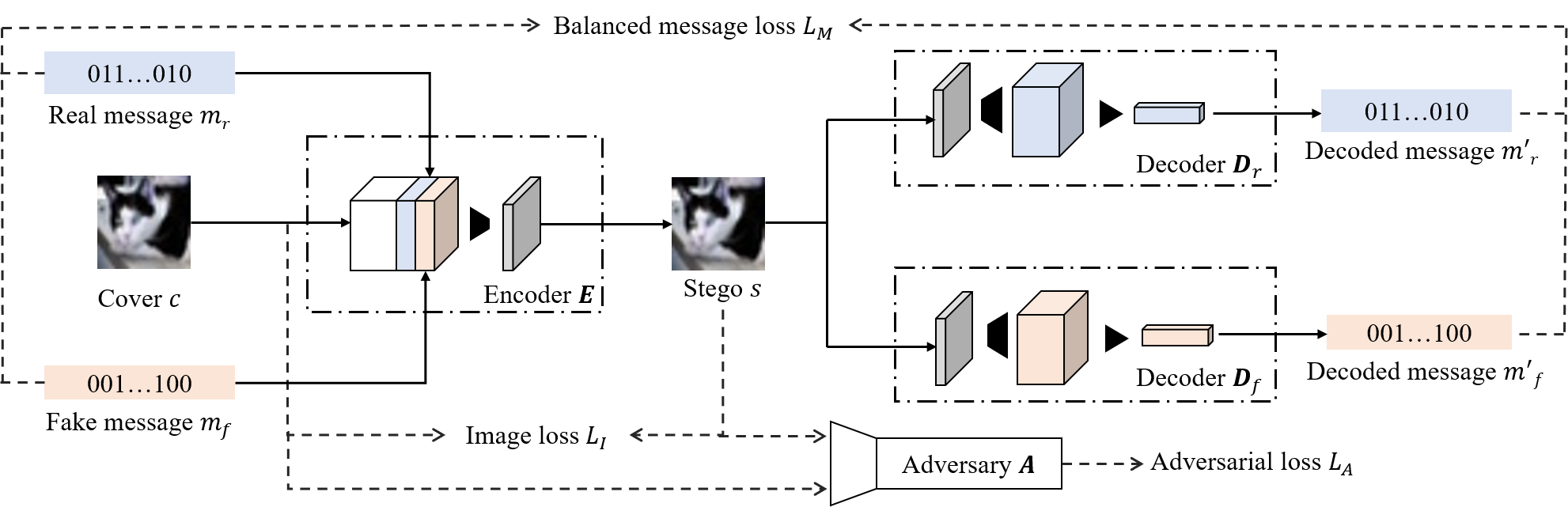}
	\caption{
		The network architecture of our scheme. The network model is implemented based on HiDDeN \cite{zhu2018hidden}, consisting of an encoder $\bm{E}$, a decoder $\bm{D_{r}}$ for the real message, another decoder $\bm{D_{f}}$ for the fake one, and an adversary $\bm{A}$. }
	\label{fig:scheme_loss}
\end{figure*}

\subsection{Feasible constructions}

 we simply sketch several feasible constructions for deniable steganography. The manual steganographic techniques, deep learning techniques, and deniable encryption can be used to build deniable steganographic schemes.  
 
 \textit{Sender-deniable steganography}. The main problem is how to design $\bm{E}_f$. One can use the XOR operation (a symmetric deniable encryption technique) combined with manual steganographic techniques to make a construction. The message can be firstly encrypted by XOR and then embedded by manual steganographic techniques. The extraction process is the reverse. The deniable property is provided by XOR.  
 
 \textit{Receiver-deniable steganography}. The main problem is how to design $\bm{D}_f$, and there are several kinds of feasible constructions: 1) the combination of the XOR operation and manual steganographic techniques can also be used for receiver-deniable steganography; 2) we can embed the real and fake messages in different locations of the cover. Then two extractors can be designed to extract the real and fake messages from different locations of the corresponding stego; and 3) we can concatenate the real and fake messages and encode them together with the cover by deep learning techniques. Then two decoders can be designed to extract the real and fake messages respectively.
 
 In this paper, we propose a receiver-deniable steganographic scheme using deep neural networks with images as covers. 

\section{A receiver-deniable image steganographic scheme}\label{sec:scheme}

\subsection{The framework}\label{sec:scheme_framework}

 In a common steganographic scheme as shown in Fig.~\ref{fig:FrameworkOfCommunication} (b), a secret message is embedded into a cover image to generate a stego image by the sender. The stego image is visually similar to the cover image. Then, the stego image is transmitted to the receiver for secure communication. Finally, the receiver extracts the secret message from the stego image. Deep steganographic methods generally have an encoder-decoder framework, with the encoder and decoder networks respectively performing the embedding and extraction operations. In addition, a steganalysis network can be incorporated into the network during the training process as an adversary module to enhance the undetectability. 
 
 Our receiver-deniable steganography scheme is adapted from deep steganographic methods. As illustrated in Fig.~\ref{fig:scheme_framework}, besides the real message $m_r$, a piece of fake massage $m_f$ is also embedded into the cover $c$ by the encoder $\bm{E}$. The generated stego $s$ will be sent to the receiver for secure communication. The fake message $m_f$ has nothing to do with the real one $m_r$ but looks like a real message. Once receiving the stego $s$, the receiver can extract the real message $m_r$ by the decoder $\bm{D}_r$. But if coerced by the adversary, the receiver can extract the fake message $m_f$ by the decoder $\bm{D}_f$ to deceive the adversary.

\subsection{The network architecture}\label{sec:architecture}

 During the training process, our scheme involves an encoder $\bm{E}$, two decoders $\bm{D}_r, \bm{D}_f$, and an adversarial discriminator $\bm{A}$, as shown in Fig.~\ref{fig:scheme_loss}. At first, each cover $c$ and two messages $m_r, m_f$ are concatenated \textit{orderly} and then encoded together by $\bm{E}$ to generate the stego $s$. Then, the stego $s$ is fed into $\bm{D}_r$ and $\bm{D}_f$ to extract $m'_r, m'_f$, respectively. $m'_r, m'_f$ are expected to be identical to $m_r, m_f$ respectively. In addition, the cover $c$ and the stego $s$ are inputted into $\bm{A}$ to estimate their classification for adversarial training.

\subsection{Loss functions}\label{sec:scheme_loss}
 
 Three kinds of loss functions are considered in our scheme, including the image loss $L_{I}$, the balanced message loss $L_{M}$, and the adversarial loss $L_{A}$.
 

 \textbf{Image loss}. It is applied to minimize the image distortion caused by data embedding. The typical mean square error is calculated as the image loss as follows,

	\begin{equation}
		L_{I}\left(c, s \right)=\frac{1}{n} \sum_{i=1}^{n}\left(c[i]-s[i]\right)^{2},
	\end{equation}

 \noindent where $c[i] $ and $s[i]$ are the $i$-th pixels in the cover $c$ and the stego $s$, and $n$ denotes the number of image pixels.

 \textbf{Balanced message loss}. It is designed to guarantee the even extraction accuracy of secret messages. There are two decoders in our scheme. Accordingly, two message loss functions are considered as follows,

	\begin{equation}\label{eq:msgloss}
		\begin{aligned}
			L_{m_r}\left(m_r, m'_r\right) &= \frac{1}{t} \sum_{i=1}^{t}\left(m_r[i]-m'_r[i]\right)^{2},\\	 
			L_{m_f}\left(m_f, m'_f\right) &= \frac{1}{t} \sum_{i=1}^{t}\left(m_f[i]-m'_f[i]\right)^{2},
		\end{aligned}
	\end{equation}
	\begin{equation}
		L_b = \left| L_{m_r}-L_{m_f} \right|.
	\end{equation}

	\begin{figure*}[!ht]
		\centering
		\includegraphics[width=0.9\linewidth]{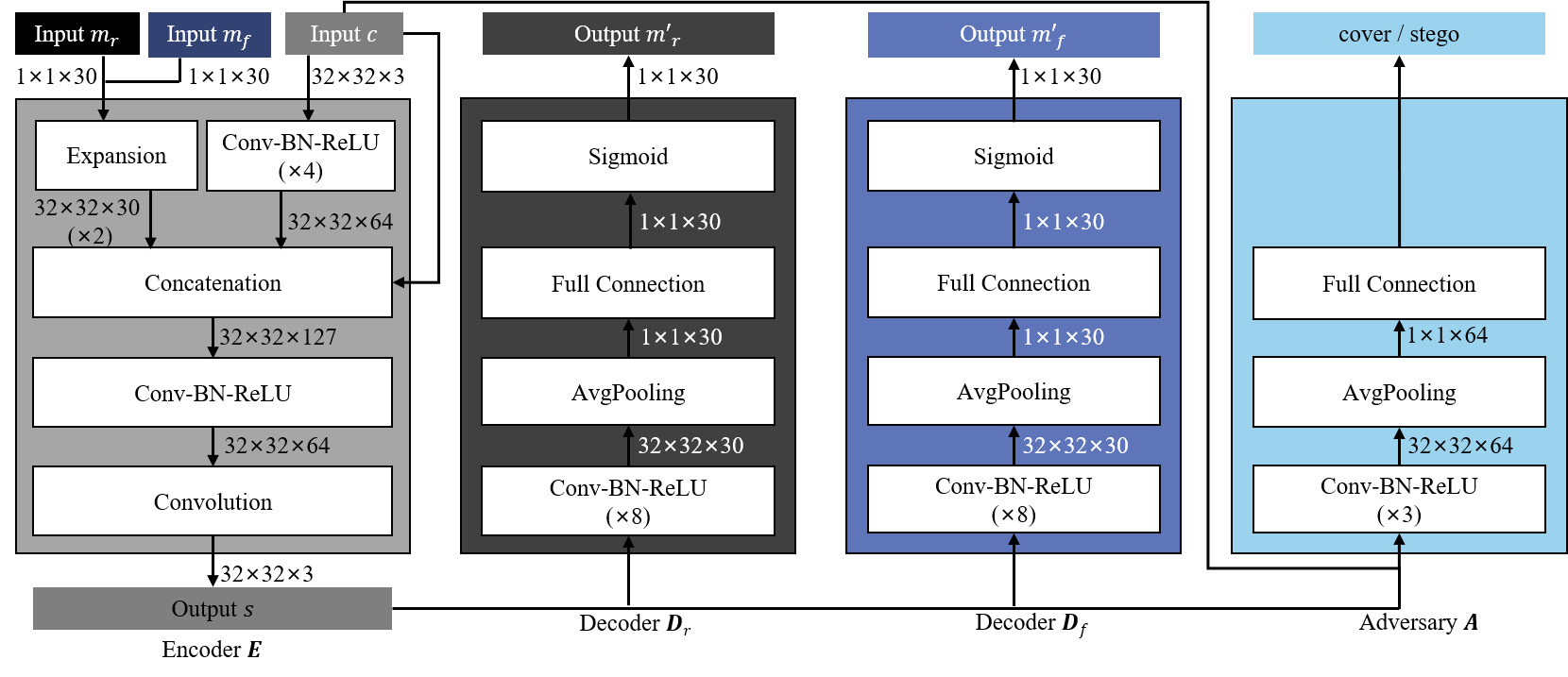}
		\caption{
			The network details of our scheme. All the 'Con-BN-ReLU' modules are the same, equipped with convolution, batch normalization, and ReLU activation in order.}
		\label{fig:scheme_details}
	\end{figure*}
 \noindent where $t$ denotes the number of message bits. In addition, a balanced performance is preferred for decoders to avoid generating an extremely bad decoder. Accordingly, we introduce a balance loss as follows,

 The final balanced message loss is a combination of the three losses above as,

	\begin{equation}\label{equ:lossmsg}
		L_{M} = \lambda_m (L_{m_r}+L_{m_f})+ \lambda_b L_b.
	\end{equation}

 \textbf{Adversarial loss}. It aims to optimize the encoder $\bm{E}$ to make the generated stego image discriminated as a cover image by the discriminator $\bm{A}$ as much as possible. As an adversarial mechanism, the encoder $\bm{E}$ and discriminator $\bm{A}$ are optimized alternately. 
 
 During the optimization of $\bm{A}$, a batch of cover images and stego images (generated by $\bm{E}$) are fed into $\bm{A}$ to estimate their labels. The binary cross-entropy loss is calculated to optimize $\bm{A}$ as follows,

	\begin{equation}		
		L_{B}\left(x,y\right)= - \left[y \log\bm{A}(x) + (1-y) \log\bm{A}\left(1-x\right)\right],
	\end{equation} 	

 \noindent where $y$ is the label of $x$ with $y=0$ if $x$ is a cover and $y=1$ if $x$ is a stego. $\bm{A}(x)$ denotes the output of $\bm{A}$ with the input $x$. 
 
 During the optimization of $\bm{E}$ as well as $\bm{D}$, a batch of cover images are fed into $\bm{E}$ to generate stego images. Then these stego images are fed into $\bm{A}$ to estimate their labels. The adversarial loss is calculated to optimize $\bm{E}$ as follows,

	\begin{equation}\label{equ:LossAdversary}
		L_{A}\left(s\right)= - \log \left(1-\bm{A}\left(s\right)\right).
	\end{equation}

 \textbf{Total loss}. Finally, the total loss function $L(\bm{w})$ can be calculated as follows,

	\begin{equation}\label{equ:AllLoss}
		L(\bm{w})=\lambda_{I} L_{I}+\lambda_{M} L_{M}+\lambda_{A} L_{A},
	\end{equation}
	
	\setlength{\tabcolsep}{2mm}{
	
		\begin{table*}[htb]
			\renewcommand{\arraystretch}{1.3}
			\caption{
				The quantitative quality of stego images when more decoders are added into the network.}
			\centering
			\begin{tabular}{c|cccccccc}
				\toprule
				Scheme                    	& $\lambda_m$	& $\lambda_b$ 	& Bits of each message & PSNR (dB)  	& SSIM    	& LPIPS	 & Bit error \\
				\midrule
				Our scheme with 2 decoders	& 1				&1				&30 &33.92      	& 0.9636  	& 0.0030  & 0.0160\\
				Our scheme with 3 decoders	& 2/3			&1/3			&20 &34.19      	& 0.9566  	& 0.0033  & 0.0134\\
				Our scheme with 4 decoders	& 2/4			&1/6			&15 &32.62      	& 0.9469  	& 0.0039  & 0.0062\\
				Our scheme with 6 decoders  & 2/6			&1/15 			&10 &31.78      	& 0.9201  	& 0.0089  & 0.0178\\
				\bottomrule
			\end{tabular}
			\label{Table:ExperimentsWithMoreDecoder}
		\end{table*}
	
	}
	
	\begin{table*}[htb]
		\renewcommand{\arraystretch}{1.3}
		\caption{
			Implementation details of different networks for deniable steganography.}
		\centering
		\begin{threeparttable}[b]
			\begin{tabular}{c|ccccccc}
				\toprule
				Network                                & Architecture  & Loss composition                & Optimizer	 & Image size       & Message length (bit) & Batch size	& Epochs \\
				\midrule
				HayesGAN \cite{hayes2017generating}	   & GAN	   & $L_{I}$, $L_{M}$, $L_{A}$	         & Adam	         & $32 \times 32$   & 100                  & 32	        & 150 \\
				DeepStego \cite{baluja2017hiding}	   & EDA	   & $L_{I}$, $L_{M}$	                 & Adam	         & $200 \times 200$ & 960,000               & 10	        & 100 \\
				HiDDeN \cite{zhu2018hidden}	           & GAN	   & $L_{I}$, $L_{M}$, $L_{A}$	         & Adam	         & $32 \times 32$   & 30                   & 12	        & 300 \\
				StegaStamp \cite{tancik2020stegastamp} & GAN	   & $L_{I}$, $L_{M}$, $L_{A}$, $L_{P}$  & Adam, RMSProp & $400 \times 400$ & 100                  & 4	        & 100 \\
				\bottomrule
			\end{tabular}
			\begin{tablenotes}
				\footnotesize
				\item `EDA' denotes the Encoder-Decoder Architecture. $L_I$, $L_M$, and $L_A$ denote the image loss, the message loss, and the adversarial loss respectively. In StegaStamp \cite{tancik2020stegastamp}, $L_{P}$ denotes the LPIPS perceptual loss \cite{zhang2018unreasonable}.
			\end{tablenotes}
		\end{threeparttable}
		\label{Table:Sensitivity1}
	\end{table*}
	
	\begin{table*}[htb]
		\renewcommand{\arraystretch}{1.3}
		\caption{
			Sensitivity of our scheme to different networks.}
		\centering
		\begin{threeparttable}[b]
			\begin{tabular}{r|ccccccc}
				\toprule
				\makecell[c]{Network}							& Time cost (s)	& Total loss & PSNR (dB) & SSIM     & LPIPS	    & Bit error\\ \midrule
				Original HayesGAN \cite{hayes2017generating} 	& {21.06} 	    & {0.0467}	 & {29.14}   & {0.6984} & {0.0738}	& {0.0013}\\
				Our scheme by HayesGAN                          & {22.11} 	    & {0.0452}	 & {28.49}   & {0.5391} & {0.0897}	& {0.0353} \\ \hline
				Original DeepStego \cite{baluja2017hiding}		& {76.69} 	    & {0.0272}	 & {33.63}   & {0.9799} & {0.0136}	& {0.2848}\\
				Our scheme by DeepStego                     	& {88.97}	    & {0.0348}	 & {31.10}   & {0.9470} & {0.0163}  & {0.2924}\\ \hline
				Original HiDDeN \cite{zhu2018hidden}			& {31.93} 	    & {0.0575}	 & {34.56}   & {0.9544} & {0.0017}  & {0.0105}\\
				Our scheme by HiDDeN                       		& {39.72}       & {0.0502}	 & {33.92}   & {0.9636} & {0.0030}  & {0.0160}\\ \hline
				Original StegaStamp \cite{tancik2020stegastamp}	& {927.18} 	    & {0.0014}	 & {36.88}   & {0.9766} & {0.0002}  & {0.0}\\
				Our scheme by StegaStamp                        & {960.96}	    & {0.0029}	 & {33.38}   & {0.9016} & {0.0010}  & {0.0}\\ \bottomrule
			\end{tabular}
			\begin{tablenotes}
				\footnotesize
				\item `Time cost' presents the average time cost per epoch. `Total loss' presents the total loss of the networks in the last epochs.
			\end{tablenotes}
		\end{threeparttable}
		\label{Table:Sensitivity2}
	\end{table*}

 \noindent where $\bm{w}$ denotes the parameters to be learned in $\bm{E}$ and $\bm{D}$, and $\lambda_{E}$, $\lambda_{D}$, and $\lambda_{A}$ are the weight coefficients that can be adjusted to make trade-offs among the three losses according to the actual demand.

\subsection{The network details} \label{sec:scheme_details}

 Here we specify the components of our scheme, including the encoder $\bm{E}$, decoders ($\bm{D}_r$, $\bm{D}_f$), and the adversarial discriminator $\bm{A}$. In fact, many excellent components in existing DNN-based steganographic methods can be utilized to construct our scheme, and we choose the structures in HiDDeN \cite{zhu2018hidden}. As illustrated in Fig. \ref{fig:scheme_details}, the secret messages $m_r$ and $m_f$ are the binary sequence with 30 random bits, the covers are color images with the size of $32 \times 32 \times 3$, and all the 'Conv-BN-ReLU' modules refer to the sequence of convolution, batch normalization, and ReLU activation \cite{glorot2011deep}. If not specified, all the convolution operations use 64 convolutional kernels with the size of $3 \times 3$, stride of 1, and padding of 1.
 
 \textbf{Encoder} $\bm{E}$. With the cover $c$ as input, an activation volume of $32 \times 32 \times 64$ is generated using 4 Conv-BN-ReLU modules. The messages $m_r$ and $m_f$ are replicated to produce two volumes with the size of $32 \times 32 \times 30$. Then, the resulting volumes and the original cover are concatenated to generate a volume of $32 \times 32 \times 127$. Subsequently, a Conv-BN-ReLU module and a normal convolution are used consecutively to produce the stego $s$.

 \textbf{Decoders} $\bm{D}$. The decoders $\bm{D}_r$ and $\bm{D}_f$ are designed to extract real and fake messages respectively. The two decoders have the same structure and are trained together, but independent of each other. At first, the stego $s$ is processed by 8 Conv-BN-ReLU modules to obtain a volume of $32 \times 32 \times 30$. Then, the messages $m_r$ and $m_f$ are produced by the global adaptive average pooling, linear full connection, and sigmoid activation operations. The global adaptive average pooling can handle stego images of arbitrary sizes, and sigmoid activation can constrain the extracted data to be in the range of 0 $\sim$ 1.

 \textbf{Adversary $\bm{A}$.} It is a discriminator with 3 Conv-BN-ReLU modules, a global adaptive average pooling, and a linear full connection. It judges whether the input image is a stego or not. The adversarial training consists of two stages that are conducted alternately. During the optimization of $\bm{A}$, the inputs are covers and stegos, and the discriminator $\bm{A}$ is optimized with the correct labels of each class. During the optimization of $\bm{E}$, the inputs are only stegos, and the encoder $\bm{E}$ is optimized with the labels of covers according to the loss function of Eq.~(\ref{equ:LossAdversary}). As a result, the generated stegos are discriminated as covers as much as possible.

\section{Experiments} \label{sec:experiments}

 This section tests the effectiveness of the proposed receiver-deniable scheme. Firstly, We present the experimental settings. Secondly, the quality of stego images and extraction accuracy of messages are evaluated to demonstrate the effectiveness of the scheme. Thirdly, the scalability of our scheme is verified by considering more decoders. Finally, the sensitivity of our scheme is tested by considering other network components. 

\subsection{Experimental settings}\label{sec:Experimental_setting}

 Here we introduce the datasets, training parameters, and evaluation metrics.

 \textbf{Datasets.} We utilize ImageNet \cite{deng2009imagenet} for training, and test the trained models on the BOSS dataset \cite{bas2011break}. All images are resized to $32 \times 32 \times 3$ before use. The secret messages are randomly generated binary sequences of 30 bits.

 \textbf{Training parameters.} Adam optimizer \cite{kingma2014adam} is used for optimization with the batch size of 12 for 300 epochs. The weight coefficients in Eq.~(\ref{equ:lossmsg}) are set as  $\lambda_m=1$ and $\lambda_b = 1$, and the weight coefficients in Eq.~(\ref{equ:AllLoss}) are set as $\lambda_{I} = 0.7$, $\lambda_{M} = 1$, and $\lambda_{A} = 0.001$ as that in HiDDeN \cite{zhu2018hidden}. 
 

 \textbf{Evaluation metrics.} The quality of stego images is evaluated by Peak Signal to Noise Ratio (PSNR), Structural Similarity (SSIM), and Learned Perceptual Image Patch Similarity (LPIPS). PSNR and SSIM are two common image quality metrics and LPIPS \cite{zhang2018unreasonable} is a network-learned perceptual similarity metric. The extraction accuracy of messages is evaluated by the averaged bit error.

 \begin{figure}
	\centering
	\subfigure[Covers]{\includegraphics[width=.27\linewidth]{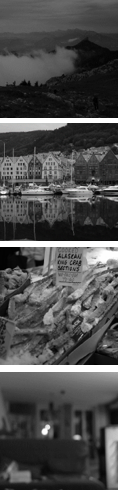}}
	\subfigure[HiDDeN\cite{zhu2018hidden}]{\includegraphics[width=.27\linewidth]{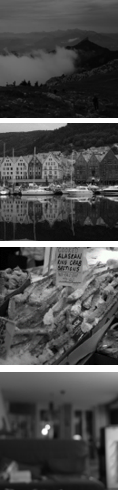}}
	\subfigure[Ours]{\includegraphics[width=.27\linewidth]{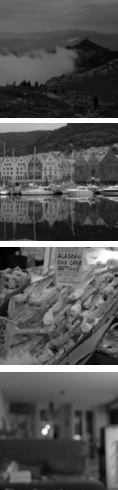}}
	\centering
	\caption{The visual effectiveness of stegos generated by HiDDeN \cite{zhu2018hidden} and our scheme.}
	\label{fig:ImgQuality}
\end{figure}

\subsection{The quality of stego images} \label{sec:ImgQuality}
 The quality of stego images is guaranteed by the image loss $L_I$. The main network of our scheme is adapted from the components in a famous steganographic network HiDDeN \cite{zhu2018hidden}. So we mainly compared our scheme with HiDDeN in the following. 

 The visual effectiveness of stego images generated by HiDDeN and our scheme is shown in Fig.~\ref{fig:ImgQuality}. For HiDDeN, the message length is set to be 60 bits. It shows that both schemes leave no evident blemishes. In addition, we compared the image qualities of the two schemes in terms of PSNR, SSIM, and LPIPS. As presented in Table~\ref{Table:ImgQuality}, our scheme performs similarly to HiDDeN.

 \setlength{\tabcolsep}{5mm}{
	\begin{table}[htb]
		\renewcommand{\arraystretch}{1.3}
		\caption{The quantitative quality of stego images.}
		\centering
		\begin{tabular}{l|ccc}
			\hline
			\makecell[c]{Scheme}				&PSNR (dB)	&SSIM		&LPIPS \\
			\hline 
			HiDDeN \cite{zhu2018hidden} 		& 35.65	    & 0.9627	& 0.0026\\	
			Our scheme 							& 35.07	    & 0.9539	& 0.0029\\
			\hline	
		\end{tabular}
		\label{Table:ImgQuality}
	\end{table} 

 }

\subsection{The extraction accuracy of message bits} \label{sec:MsgAccuracy} 
 Different from the original HiDDeN, we added a sigmoid layer to the decoders and a balance loss $L_{b}$ to balance the extraction accuracy between the real and fake messages. As listed in Table~\ref{Table:MsgAcu}, our scheme outperforms HiDDeN slightly in terms of bit error. In addition, the sigmoid layer decreases the extraction error, and the balance loss is also effective. The results in Table~\ref{Table:MsgAcu} are averaged from 1000 test images.
 
 \begin{table}[htb]
	\renewcommand{\arraystretch}{1.3}
	\caption{The extraction bit errors.}
	\centering
	\begin{tabular}{l|cc}
		\hline
		\makecell[c]{Scheme}				& $m_r$			& $m_f$		\\
		\hline
		HiDDeN \cite{zhu2018hidden} 		& 0.0334	    & N/A 		\\
		Our scheme without balance loss		& 0.0385	    & 0.0047 	\\
		Our scheme without sigmoid layer	& 0.0325	    & 0.0186 	\\
		Our scheme with all components	    & 0.0293	    & 0.0173 	\\
		\hline
	\end{tabular}
	\label{Table:MsgAcu}
\end{table}

\subsection{Scalability with more decoders} \label{sec:experiments_freedom}

 A natural extension of our scheme is to embed more than two pieces of messages in a cover. Accordingly, more decoders need to be incorporated into the network for message extraction. In this design, the multiple messages are expanded and concatenated with the cover together, the increased decoders can have the same structure as that in Fig.~\ref{fig:scheme_details}, and the balance loss $L_{b}$ needs to be adjusted as follows,

	\begin{equation}
		L_{b} = \sum_{i=1, j=1}^{n}\left|L_{m_i}-L_{m_j}\right|, i \neq j,
	\end{equation}

\noindent where $L_{m_i}$ can be calculated as that in Eq.~(\ref{eq:msgloss}). With more decoders, the parameters in Eq.~(\ref{equ:lossmsg}) need adjustments to balance the weights between the image quality and extraction accuracy of message bits. In addition, with more decoders, and the bits of each piece of message are decreased to remove the effect of payload. As listed in Table~\ref{Table:ExperimentsWithMoreDecoder}, the proposed scheme holds similar results with different numbers of decoders.

\subsection{Sensitivity to different network architectures}\label{sec:experiments_backbones}

 In the experiments above our scheme is implemented using the network in HiDDeN \cite{zhu2018hidden}. In fact, our scheme can be implemented using the architectures in some other steganographic networks, such as DeepStego \cite{baluja2017hiding}, HayesGAN \cite{hayes2017generating}, and StegaStamp \cite{tancik2020stegastamp}. The implementation details are listed in Table~\ref{Table:Sensitivity1}, and the results are listed in Table~\ref{Table:Sensitivity2}. It shows that our scheme achieves comparable performance with the corresponding original schemes. It demonstrates that our scheme is not sensitive to the network architectures and can be applied to different steganographic networks.

\section{Conclusion}\label{sec:conclusion}

 This paper considers the coercing problem in steganography for the first time, and gives the definitions and feasible constructions in deniable steganography. We proposed a receiver-deniable image steganographic scheme using deep neural networks. The scheme can be adapted from the existing DNN-based steganographic networks with just a little adjustment in the design of loss function. The experiments demonstrate the scalability and sensitivity of our scheme. In the future, we will try to design the sender-deniable steganography schemes and the schemes that take a secret key to control the extraction of messages. In addition, our idea can be extended to construct DNN-based deniable image encryption.

\section*{Acknowledgment}
\label{sec:acknowledgment}

This work is supported in part by the National Key Research and Development Plan of China under Grant 2020YFB1005600, in part by the National Natural Science Foundation of China under grant numbers 62122032, 62172233, 62102189, U1936118, 61825203, U1736203, and 61732021, in part by the Major Program of Guangdong Basic and Applied Research Project under Grant 2019B030302008, in part by Six Peak Talent project of Jiangsu Province (R2016L13), Qinglan Project of Jiangsu Province, and "333" project of Jiangsu Province, in part by the National Joint Engineering Research Center for Network Security Detection and Protection Technology, in part by the Priority Academic Program Development of Jiangsu Higher Education Institutions (PAPD) fund, in part by the Collaborative Innovation Center of Atmospheric Environment and Equipment Technology (CICAEET) fund, China. Zhihua Xia is supported by BK21+ program from the Ministry of Education of Korea.

\bibliographystyle{IEEEtran}
\bibliography{IEEEabrv, DS}

\begin{IEEEbiography}[{\includegraphics[width=1in,height=1.25in,clip,keepaspectratio]{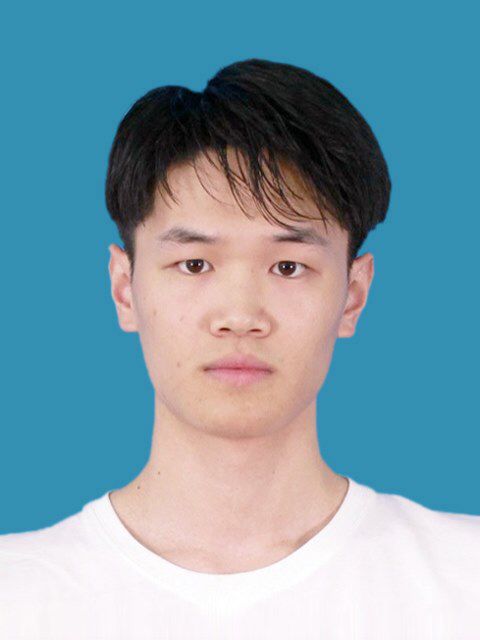}}]{Yong Xu} received his BE degree in Nanjing University of Information Science \& Technology in 2020. He is currently pursuing master degree in School of Computer and Software in Nanjing University of Information Science \& Technology. His research interests include data hiding and information forensics.
\end{IEEEbiography}

\begin{IEEEbiography}[{\includegraphics[width=1in,height=1.25in,clip,keepaspectratio]{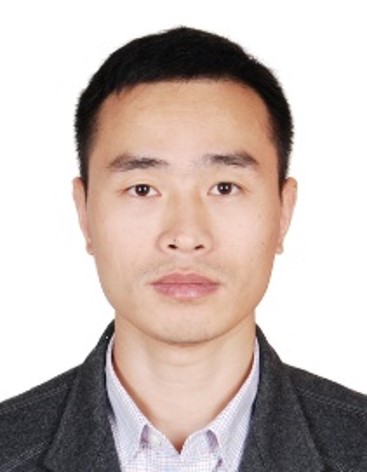}}]{Zhihua Xia} received a BS degree in Hunan City University, China and PhD degree in computer science and technology from Hunan University, China, in 2006 and 2011, respectively. He works as an associate professor in the School of Computer and Software, Nanjing University of Information Science and Technology. His research interests include digital forensic and encrypted image processing. He is a member of the IEEE from 1 March 2014.
\end{IEEEbiography}

\begin{IEEEbiography}[{\includegraphics[width=1in,height=1.25in,clip,keepaspectratio]{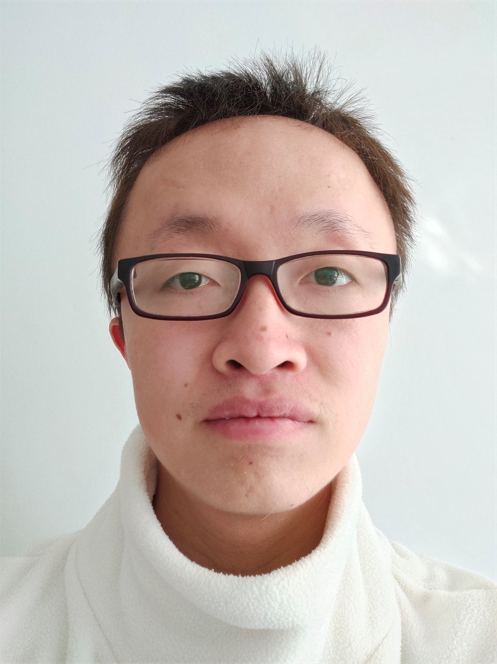}}]{Zichi Wang} received the BS degree in electronics and information engineering from Shanghai University, China, in 2014, and received the MS degree in signal and information processing in 2017, the PhD degree in information and communication engineering from the same university in 2020. His research interests include steganography, and steganalysis. He has published over 40 papers in these areas. E-mail: wangzichi@shu.edu.cn	
\end{IEEEbiography}

\begin{IEEEbiography}[{\includegraphics[width=1in,height=1.25in,clip,keepaspectratio]{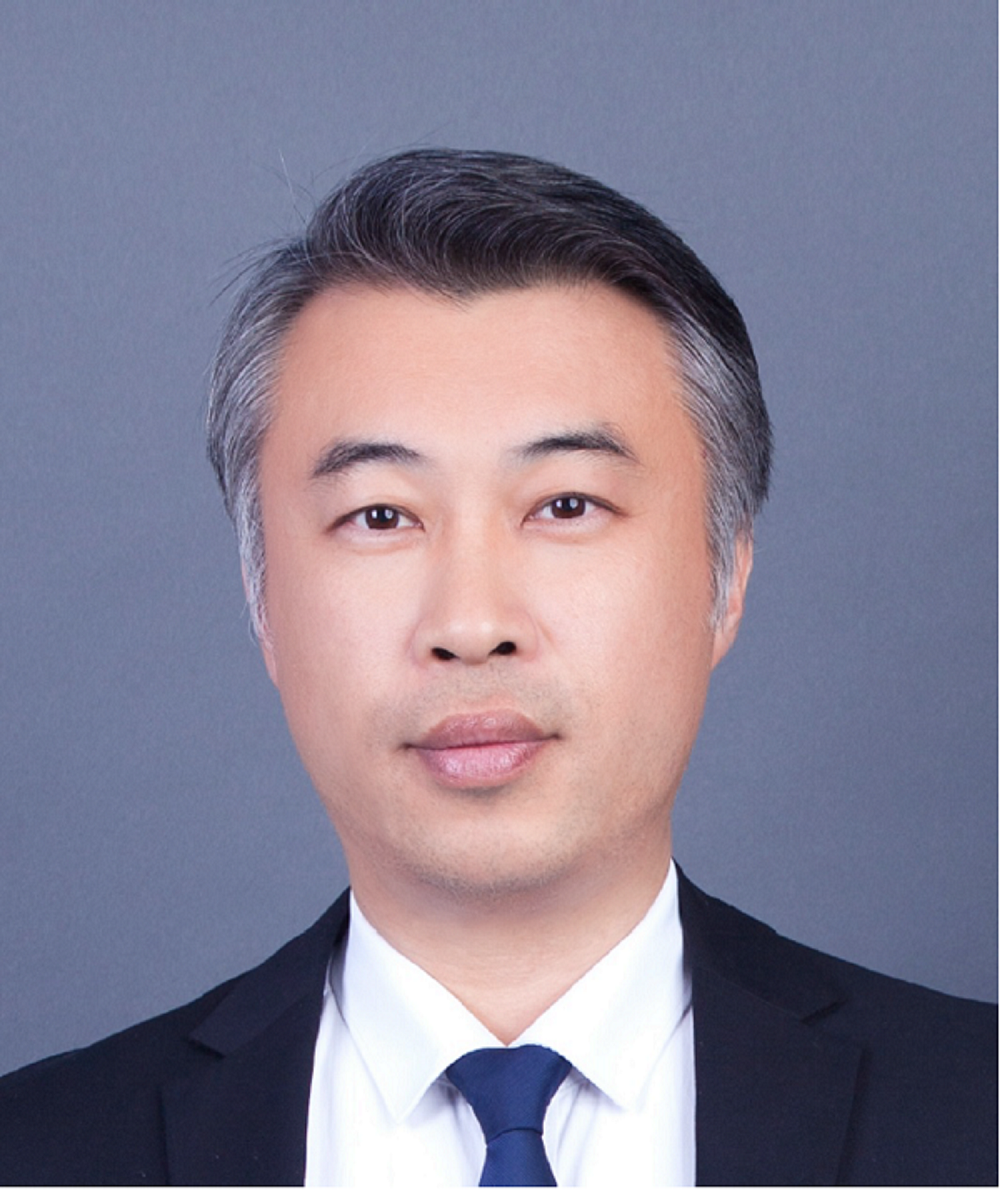}}]{Xingpeng Zhang} received B.S. from Jilin University, China, in 1995, and the M.S. and Ph.D. from Shanghai University, in 2001 and 2004, respectively. Since 2004, he has been with the faculty of the School of Communication and Information Engineering, Shanghai University, where he is currently a fulltime Professor. He is also with the faculty of the School of Computer Science, Fudan University. He was with The State University of New York at Binghamton as a Visiting Scholar from 2010 to 2011, and also with Konstanz University as an experienced Researcher, sponsored by the Alexander von Humboldt Foundation from 2011 to 2012. His research interests include multimedia security, image processing, and digital forensics. He has published over 200 research papers. He was an Associate Editor for IEEE Transactions on Information Forensics and Security from 2014 to 2017. Email: xzhang@shu.edu.cn		
\end{IEEEbiography}

\begin{IEEEbiography}[{\includegraphics[width=1in,height=1.25in,clip,keepaspectratio]{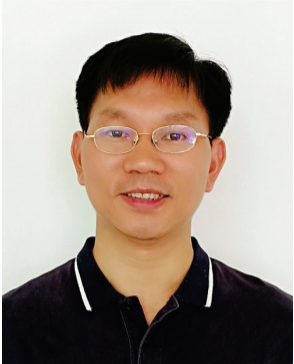}}]{Jian Weng}   
	received the Ph.D. degree in computer science and engineering from Shanghai Jiao Tong University, Shanghai, China, in 2008. He is currently a Professor and the Dean with the College of Information Science and Technology, Jinan University, Guangzhou, China. His research interests include public key cryptography, cloud security, and block-chain. He was the PC Co-Chairs or PC Member for more than 30 international conferences. He also serves as an Associate Editor for the IEEE Transactions on Vehicular Technology. 
\end{IEEEbiography}

\end{document}